\documentclass[12pt,preprint]{aastex}

\def\eg{{\it e.g.}\ } 
 
\def\ie{{\it i.e.}\ }
\def\bb{{\bf B}}
\def\bj{{\bf J}}
\def\rr{{\bf r}}
\def\be{\begin{equation}}
\def\ee{\end{equation}}

\shorttitle{A Universal Law for Solar-Wind Turbulence at Electron Scales}
\shortauthors{Meyrand and Galtier}

\begin{document}

\title{A Universal Law for Solar-Wind Turbulence at Electron Scales}
\author{R. Meyrand and S. Galtier\altaffilmark{1}}
\affil{Univ Paris-Sud, Institut d'Astrophysique Spatiale, b\^atiment 121, F-91405 Orsay, France}
\email{rmeyrand@ias.fr,sebastien.galtier@ias.fr}
\altaffiltext{1}{also at Institut universitaire de France}

\begin{abstract}
The interplanetary magnetic fluctuation spectrum obeys a Kolmogorovian power law at scales above 
the proton inertial length and gyroradius which is well regarded as an inertial range. Below these scales 
a power law index around $-2.5$ is often measured and associated to nonlinear dispersive processes. 
Recent observations reveal a third region at scales below the electron inertial length. This region is 
characterized by a steeper spectrum that some refer to it as the dissipation range. We investigate this 
range of scales in the electron magnetohydrodynamic approximation and derive an exact and universal 
law for a third-order structure function. This law can predict a magnetic fluctuation spectrum with an index 
of $-11/3$ which is in agreement with the observed spectrum at the smallest scales. We conclude on the 
possible existence of a third turbulence regime in the solar wind instead of a dissipation range as recently 
postulated. 
\end{abstract}


%%%%%%%%%%%%%%%
\section{Introduction}

Turbulence plays a central role in a wide range of astrophysical plasmas. Examples are given by the 
solar wind \citep{mat99}, the interstellar \citep{scalo}, galactic and even intergalactic media 
\citep{Govoni}. In the solar wind, turbulence evolves freely and is not perturbed by \textit{in situ} 
diagnostics, therefore it provides an ideal laboratory for studying high Reynolds number plasma 
turbulence. This unique situation allows us to investigate for example the origin of anisotropy 
\citep[see \eg][]{klein,gal00,alex07,big08}, to evaluate the mean energy dissipation rate 
\citep{macbride,carb}, to detect multiscale intermittency \citep{Kiyani09}, or to analyze different regimes 
of turbulence characterized by a steepening of the magnetic field fluctuations spectrum with a power law 
index going from $-5/3$, at frequencies lower than $1$Hz, to indices lying around $-2.5$ at higher 
frequencies \citep[see \eg][]{Smith06}. 

The spectral break near 1Hz has been a subject for intensive studies and controversies in the last decades. 
It was first interpreted as the onset of dissipation caused for example by kinetic Alfv\'en wave damping 
\citep{Leamon}. Then, it was demonstrated that the wave damping rate usually increases very strongly with 
wavenumbers and should lead to a strong cutoff in the power spectra rather than a steepened power law
 \citep{li}. In the meantime, there are some indications that the fluctuations are accompanied by a bias of the 
polarization suggesting the presence of right-hand polarized, outward propagating waves \citep{Golstein94}. 
Also it was proposed \citep{Stawicki01} that Alfv\'en -- left circularly polarized -- fluctuations are suppressed 
by proton cyclotron damping and that the high frequency power law spectra are likely to consist of whistler 
fluctuations \citep{matt}. It is currently believed that the steepening of the spectra at 1Hz is mainly due to 
non-linear dispersive processes  that range from kinetic Alfv\'en waves \citep{hasegawa,howes}, 
electromagnetic ion-cyclotron Alfv\'en waves \citep{wu}, or/and electron whistler waves 
\citep{Ghosh96,gal06,gal07,nat10} in the framework of Hall magnetohydrodynamics (MHD) or simply 
electron MHD. 

The most recent solar wind observations made with the high resolution magnetic field data of the 
Cluster spacecraft \citep{sahraoui09,alexandrova} reveal the presence of a third region -- called 
dissipation range -- at scales smaller than $d_e$ and characterized by even steeper magnetic 
fluctuations spectra with a power law index around $-3.8$. These spectra observed only on half a decade 
are interpreted as either a power law \citep{sahraoui09} or an exponential law \citep{alexandrova}. 
Although the theoretical interpretation of such a regime is still open \citep{matt}, a recent theoretical 
analysis shows that a kinetic Alfv\'en wave cascade subject to collisionless damping cannot reach 
electron scales in the solar wind at 1 AU \citep{podesta10}. The direct consequence is that the spectra 
observed must be supported by another type of wave modes. It is noteworthy that this new regime at 
electron scales gives rise to the same controversy as the steepening found two decades ago around 1Hz
which brings up naturally the following question: Have we really found the dissipation scale of the solar 
wind plasma or is it the onset of a new turbulence regime?

In this article, we investigate the turbulence regime at scales smaller than the electron inertial 
length $d_e$ through the electron MHD approximation. The assumption of homogeneity and 
isotropy will be made to derive an exact and universal law for third-order structure functions. 
We show that this law corresponds to a magnetic fluctuations isotropic spectrum in $k^{-11/3}$
compatible with the solar wind measurements. Although the assumption of isotropy is in apparent 
contradiction with the observations, it is claimed that the method used is a powerful way to have a 
first estimate of the anisotropic spectrum. Indeed, the main source of anisotropy is the presence of 
a large scale magnetic field which reduces the nonlinear transfer along its direction. Then, the most 
relevant spectral scaling is the transverse one for which the spectral index corresponds to the isotropic 
case if arguments based on the critical balance condition are used. It is only in the asymptotic limit of wave 
turbulence -- for which anisotropy is strong -- that the spectral index for transverse fluctuations is (slightly) 
modified (see the review by \cite{galtier09}).  
Finally, we conclude the paper on the possible existence of a third inertial range for solar wind turbulence 
instead of a dissipation range as recently postulated.

%%%%%%%%%%%%%%%
\section{Electron magnetohydrodynamics}

Electron MHD provides a fluid description of the plasma behavior on length scales smaller than the ion 
inertial length $d_i$ and on time scales of the order of, or shorter than, the ion cyclotron period \citep{kinsep}. 
In this case ions do not have time to respond because of their heavy mass and merely provide a neutralizing 
background. Then, the plasma dynamics is governed by electron flows and their self-consistent magnetic field. 
This model has attracted a lot of interest because of its potential applications in fast switches, Z--pinches, 
impulsive magnetospheric/solar corona reconnection and ionospheric phenomena 
\citep[see \eg][]{bhatt,cha}. 

The inviscid three-dimensional electron MHD equations can be written in SI as \citep{biskamp96}
\be
\partial_t (1 - d_e^2 \Delta) \bb = - d_i \nabla \times [ \bj \times (1 - d_e^2 \Delta) \bb ] \, , 
\label{e1}
\ee
where $\bb$ is the magnetic field normalized to a velocity ($\bb \to \sqrt{\mu_0 \rho} \, \bb$) 
and $\bj=\nabla \times \bb$ is the normalized current density. Under the limit of electron MHD we 
remind that the current density is proportional to the electron velocity. This equation has two invariants 
\citep{biskamp99} which are the total energy
\be
E = {1 \over 2} \int (B^2 + d_e^2 J^2) d {\bf x} \, , 
\ee
and the generalized helicity
\be
H = \int ({\bf A} - d_e^2 \bj) \cdot (\bb - d_e^2 \Delta \bb) d {\bf x} \, ,
\ee
with ${\bf A}$ the normalized magnetic potential.

Equation (\ref{e1}) is often used when $d_e \to 0$, namely for scales between $d_i$ and $d_e$, 
for which a $k^{-7/3}$--isotropic magnetic energy spectrum is found numerically and heuristically 
\citep{biskamp96,biskamp99,dast1}. This result is compatible with a rigorous derivation of a universal 
law for third-order correlation tensors \citep{galtier08a,galtier08b}. A steeper magnetic spectrum in 
$k^{-11/3}$ may also be found when the kinetic energy overtakes the magnetic energy \citep{gal07}. 
Such a situation -- generally not discussed in the literature -- can only be observed when the full Hall 
MHD system is considered. Note that this $-11/3$ power law index, valid for length scales larger than 
$d_e$, has a different origin from the one derived in the present paper which is applicable for scales
shorter than $d_e$. 

The behavior at scales shorter than $d_e$ has attracted much less attention \citep{biskamp96,dast2}. 
This regime corresponds to the limit $d_e \to +\infty$ for which we have 
$\partial_t \Delta \bb=- d_i\nabla \times [ \bj \times \Delta \bb ]$, 
or equivalently with the relation $\Delta \bb = - \nabla \times \bj$,
\be
{1 \over d_i} \partial_t \bj= - \bj \times (\nabla \times \bj) - \nabla \Phi \, , 
\label{e2}
\ee
where $\Phi$ is an unknown function. The second term in the right hand side may be seen as 
a gauge; actually, an analysis performed directly on the generalized Ohm's law shows that it 
corresponds to an electron pressure. Note that the form of equation (\ref{e2}) is well adapted to the 
problem under consideration since we are going to assume isotropy which means we will not consider 
any background (large scale) magnetic field ${\bf B_0}$.

%%%%%%%%%%%%%%%
\section{Universal law for $r < d_e$}

In the following, we shall derive an exact and universal law for third-order structure functions for 
homogeneous three-dimensional isotropic electron MHD turbulence at scales smaller than $d_e$ and 
discuss the implications in terms of magnetic fluctuations spectrum. After simple manipulations on 
equation (\ref{e2}) we obtain for the $i$th--component 
\be
{1 \over d_i} \partial_t J_i= J_\ell \partial_\ell J_i - \partial_i (\Phi + J^2/2) \, , 
\label{e3}
\ee
where the Einstein notation is used. Note that we also have $\partial_\ell J_\ell = 0$. The second 
term in the right hand side is similar to a pressure whereas the first term exhibits a sign of difference 
with the usual advection term encounters in Navier-Stokes equations. Actually, the Navier-Stokes 
equations may be recovered when the electron velocity is used instead of the current density or 
when the generalized Ohm's law is directly used. We made the choice to use the well-known 
electron MHD equations (\ref{e1}) and  (\ref{e3}) mainly because we shall compare eventually the 
new prediction with the previous one for scales larger than $d_e$ \citep{galtier08a}.

It is straightforward to derive a universal law for an homogeneous and isotropic electron MHD 
turbulence. First we introduce the second-order correlation tensor 
\be
R_{ij}(\rr) \equiv \langle J_i({\bf x}) J_j({\bf x^\prime}) \rangle = \langle J_i J_j^\prime \rangle \, , 
\ee
where ${\bf x^\prime}={\bf x} + \rr$. We obtain
\begin{eqnarray}
{1 \over d_i} \partial_t R_{ij}(\rr) &=& \langle J_i J^{\prime}_\ell \partial^{\prime}_\ell J_j^{\prime} \rangle
- \langle J_i \partial^{\prime}_j (\Phi^\prime + {J^\prime}^2/2) \rangle \\
&+& \langle J^{\prime}_j J_\ell \partial_\ell J_i \rangle 
- \langle J_j^\prime \partial_i (\Phi + J^2/2) \rangle \, . \nonumber
\end{eqnarray}
After simple manipulations where we use the homogeneity and the divergence free condition
we get
\begin{eqnarray}
{1 \over d_i} \partial_t R_{ij} &=& \partial_{r_\ell} (\langle J_i J^{\prime}_\ell J_j^{\prime} \rangle
- \langle J^{\prime}_j J_\ell J_i \rangle ) \, .
\end{eqnarray}
Note that the pressure-type contributions removed as usual for isotropic turbulence \citep{batch}. 
When the diagonal part of the energy tensor is only retained we have
\begin{eqnarray}
{1 \over d_i} \partial_t R_{ii} &=& -2 \partial_{r_\ell} \langle J^{\prime}_i J_\ell J_i \rangle \\
&=& -2 \nabla \cdot \langle \bj (J_i J^{\prime}_i) \rangle \, .
\label{ez}
\end{eqnarray}
At this level of analysis it is necessary to say a word about the small scale dissipation and large 
scale forcing terms which have been neglected so far. The dissipation is a linear term which is 
seen as a sink for the energy. Since we are interested in a universal behavior of turbulence we are in 
a situation where the scales considered are supposed to be much larger than the dissipation scales: 
in other words, we are deep inside the inertial range where the dissipation has no effect. The forcing 
term is assumed to be at the largest scales and acts as a constant source of energy for the system. 
Formally, the introduction of a small scale dissipation ${\cal D}$ and a large scale force ${\cal F}$
leads to the expression 
\begin{eqnarray}
{1 \over 2} d_e^2 \partial_t R_{ii} &=& - d_i d_e^2 \nabla \cdot \langle \bj (J_i J^{\prime}_i) \rangle 
+ {\cal F} + {\cal D} \, .
\label{ezz}
\end{eqnarray}

An exact relation may be derived for third-order structure functions by assuming the following 
assumptions specific to fully developed turbulence \citep{K41,frisch,PP98}. First, we take the long time 
limit for which a stationary state is reached with a finite mean energy dissipation rate per unit 
mass. Second, we consider the infinite magnetic Reynolds number limit for which the 
mean energy dissipation rate per unit mass tends to a finite positive limit, $\varepsilon^J$ 
\citep[see \eg][]{biskamp96}. By noting that 
\be
R_{ii} = \langle J^2 \rangle - {1 \over 2} \langle \delta J_i \delta J_i \rangle \, , 
\ee
where $\delta \bj \equiv \bj({\bf x}+\rr) - \bj({\bf x})$, we obtain
\be
{1 \over 2} d_e^2 \partial_t R_{ii} = \partial_t \langle {1 \over 2} d_e^2 J^2 \rangle
- {1 \over 4} d_e \partial_t \langle \delta J_i \delta J_i \rangle \, ,
\label{diss}
\ee
where the first term in the right hand side is the time variation of energy. Therefore, in the stationary 
state both terms in the right hand side are equal to zero. Since dissipation effects are negligible we
only have to include the mean energy injection rate per unit mass $\varepsilon^J$. It is important to 
remind that the energy (an inviscid invariant) is built directly from the current density hence the name 
$\varepsilon^J$. As it will be shown below, this remark turns out to be fundamental for the prediction 
of the magnetic fluctuations spectrum. The insertion of the previous statements into (\ref{ezz}) leads to 
\be 
d_i d_e^2 \nabla \cdot \langle \bj (J_i J^{\prime}_i) \rangle = \varepsilon^J \, . 
\ee
The introduction of structure functions for the current density gives eventually\footnote{It is also possible 
to consider a system without external forcing \citep{landau}. In this case, we have to deal with the decay 
problem for which the time derivative of the energy is equal (up to a sign) to the mean energy dissipation 
rate per unit mass $\varepsilon^J$. It is still possible to assume the time independence of the second 
term in the right hand side of equation (\ref{diss}) and to finally recover the same relation as (\ref{xyz}).}
\be
d_i d_e^2 \nabla \cdot \langle \delta \bj (\delta \bj)^2 \rangle = 4 \varepsilon^J \, . 
\label{xyz}
\ee

An integration of (\ref{xyz}) over a full sphere of radius $r$ (since isotropy is assumed) and the 
application of the divergence theorem give finally the universal and exact law for three-dimensional 
homogeneous isotropic electron MHD turbulence for scales smaller than $d_e$; it writes
\be 
d_i d_e^2 \langle \delta J_L (\delta \bj)^2 \rangle = {4 \over 3} \varepsilon^J r \, ,
\label{ef}
\ee
where $L$ means the longitudinal component of the vector, \ie the one along the direction 
$\rr$. Note the positive sign in the right hand side which is compatible with the negative sign in 
front of the nonlinear term of equation (\ref{e2}). The most remarkable aspect of this law is that it 
not only provides a linear scaling for third-order structure functions within the inertial range of length 
scales, but it also fixes the value of the numerical factor appearing in front of the scaling relation.

%%%%%%%%%%%%%%%
\section{Extension to $r>d_e$}

For scales larger than $d_e$ (but still smaller than $d_i$) the universal law for three-dimensional 
isotropic electron MHD turbulence takes the form \citep{galtier08a} 
\be 
d_i \langle [(\bj \times \bb) \times \bb^\prime ]_L \rangle = - {1 \over 3} \varepsilon^M r \, ,
\label{f10}
\ee
where $L$ still means the longitudinal component. This law may also be written as \citep{galtier09z}
\be 
d_i \langle [(\overline{\bj \times \bb}) \times \delta \bb]_L  \rangle = - {1 \over 3} \varepsilon^M r \, ,
\label{f10b}
\ee
where $\overline{\bf X} \equiv ({\bf X} + {\bf X^\prime})/2$. Both universal laws (\ref{ef}) and (\ref{f10b}) 
may be gathered by noting that in this case the forcing scale is pushed at scales much larger than $d_e$.
One needs to consider the following expression in the stationary state and in the inertial range
\be
{1 \over 2} (d_e^2 \partial_t R_{ii} + \partial_t {\tilde R}_{ii}) = {\cal NL} + {\cal F} + {\cal D} 
= {\cal NL} + \varepsilon^T \, ,
\ee
where ${\tilde R}_{ij}=\langle B_i B^\prime_j \rangle$, ${\cal NL}$ is the nonlinear contribution and 
$\varepsilon^T$ is the mean total energy injection rate per unit mass. Then, one obtains the general 
law for three-dimensional isotropic electron MHD turbulence (with 
$r<d_i$)
\be 
4 d_i \langle [(\overline{\bj \times \bb}) \times \delta \bb]_L \rangle 
- d_i d_e^2 \langle (\delta \bj)^2 \delta J_L \rangle = - {4 \over 3} \varepsilon^T r \, .
\label{f11}
\ee
This universal law conserves the linearity in $r$ but emphasizes the role of each term according 
to the scale considered. The importance of each term is given by the current density which involves 
a derivative: at small scales (scales smaller than $d_e$) the contribution of the current density 
will be more pronounced than at large scales (scales greater than $d_e$).

%%%%%%%%%%%%%%%
\section{Magnetic fluctuations spectrum}

The universal law (\ref{ef}) gives a precise description of electron MHD turbulence at scales
smaller than $d_e$. It also provides the possibility to predict the form of the magnetic fluctuations 
spectrum and then a comparison with observations.  Dimensionally relation (\ref{ef}) corresponds
to 
\be
d_i d_e^2 J^3 \sim \varepsilon^J r \, ,
\ee
which means an energy spectrum in 
\be
E^J(k) \sim \left({\varepsilon^J \over d_id_e^2}\right)^{2/3} k^{-5/3} \, , 
\ee
compatible with the dynamical equation (\ref{e3}) and direct numerical simulations \citep{biskamp96}.
Since the current density and the magnetic field satisfy the equation $\bj = \nabla \times \bb$, we 
obtain the scaling relation 
\be
d_i d_e^2 {B^3 \over r^3} \sim \varepsilon^J r \, .
\ee
Then, the corresponding magnetic fluctuations power spectrum scales as 
\be
B^2(k) \sim \left({\varepsilon^J \over d_id_e^2}\right)^{2/3} k^{-11/3} \, . 
\label{magnet}
\ee
This magnetic spectrum is significantly steeper than the energy spectrum. It is 
compatible with solar wind measurements where the magnetic field fluctuations are generally 
used to investigate turbulence at small scales, \ie for frequencies $f$ higher than $1$Hz. 
A power law around $f^{-3.8}$ has been reported \citep{sahraoui09} which is significantly 
steeper than what it is found at smaller frequencies where we have a spectrum around $f^{-2.5}$. 
Note that the linear law in $r$ (\ref{f11}) corresponds in fact to a double-scaling law 
for the magnetic fluctuations spectrum in $k^{-7/3}$ and $k^{-11/3}$ for, respectively, length 
scales larger and shorter than $d_e$. If we follow the Taylor frozen-in-flow hypothesis -- which is 
questionable at this length scales -- then we arrive at the conclusion that the general law (\ref{f11}), 
although its relative simplicity with the assumptions of isotropy and incompressibility, is in good 
agreement with the observations.

%%%%%%%%%%%%%%%
\section{Conclusion}
The turbulence regime at scales smaller than the electron inertial length $d_e$ has been investigated 
through the approximation of electron MHD. A new universal and exact law has been established 
in terms of structure functions for the current density. This law leads to the prediction of a $k^{-11/3}$ 
power law spectrum for the magnetic field fluctuations compatible with the most recent observations
made with Cluster. It is proposed that electron MHD turbulence provides a valuable first order 
approximation for the solar wind dynamics in particular below the length scale $d_e$. It also provides 
the first prediction for the magnetic fluctuations spectrum at these length scales. The possibility to get 
a turbulence regime at electron scales questions the origin of dissipation in the solar wind and more generally
in space plasmas. In previous 
analyses \citep{sahraoui09,alexandrova} it was suggested that the range of scales where the heating 
occurs was discovered but it was also confessed that the characteristics of turbulence in the vicinity of 
the kinetic plasma scales are not well known neither experimentally nor theoretically and are a matter of 
debate \citep[see also][]{podesta10}. It is believed that the present theoretical prediction will help 
significantly in such a debate.

%%%%%%%%%%%%%%%
\acknowledgments

Financial support from Institut universitaire de France is gratefully acknowledged. 

%%%%%%%%%%%%%%%
\appendix

\section{Extension to 2D or slab turbulence}

We shall discuss now the extension of the exact prediction (\ref{ef}) to the non isotropic case. The simplest 
situation is when a background magnetic field ${\bf B_0}$ is applied to the plasma flow for which the dynamics 
becomes statistically axisymmetric. We will adopt the view point of \citet{macbride} and assume by simplicity 
that electron MHD turbulence is either 2D or slab. If we define the energy flux vector ${\bf F}$ as (see relation 
(\ref{xyz})) 
\be
{\bf F} \equiv d_i d_e^2 \nabla \cdot \langle \delta \bj (\delta \bj)^2 \rangle \, , 
\ee
then we obtain the general formulation for isotropic electron MHD turbulence 
\be 
F(r) = {4 \over D} \varepsilon^J r \, ,
\label{zzz}
\ee
where $D$ is the space dimension. 2D turbulence means that ${\bf F}$ and ${\bf B_0}$ are perpendicular 
whereas they are parallel for slab (or 1D) turbulence. Under these simplifications it is straightforward to derive 
the following predictions 
\be
F^{2D}(r) = 2 \varepsilon^J V_{SW} \tau \sin \theta \, ,
\ee
and
\be
F^{slab}(r) = 4 \varepsilon^J V_{SW} \tau \cos \theta \, ,
\ee
where $\tau$ is the time, and $\theta$ is the angle between the mean magnetic field and the solar wind 
velocity $V_{SW}$ directions. 
Note that to obtain these predictions the Taylor hypothesis has been used \citep{macbride}. It is believed that 
such relations could be useful for analyzing solar wind turbulence and to evaluate the perpendicular and parallel 
mean energy dissipation rates per unit mass. Note that a more sophisticated approach has been recently used to 
describe axisymmetric electron MHD turbulence for length scales larger than $d_e$ \citep{galtier09z}. Under the assumption of critical balance it leads to a vectorial relation for the energy flux with a dependence in both $r$ and 
$\theta$.

%%%%%%%%%%%%%%%

\end{document}